fig.
\documentclass[epj]{svjour}
\usepackage{graphics}
\def\beg{\begin{equation}}
\def\ee{\end{equation}}

\begin{document}
\def\nt{{\em Nature }}
\def\prl{{\em Phys. Rev. Lett. }}
\def\prc{{\em Phys. Rev. C }}
\def\prd{{\em Phys. Rev. D }}
\def\jap{{\em J. Appl. Phys. }}
\def\ajp{{\em Am. J. Phys. }}
\def\nima{{\em Nucl. Instr. and Meth. Phys. A }}
\def\epl{{\em Europhys. Lett. }}
\def\npa{{\em Nucl. Phys. A }}
\def\npb{{\em Nucl. Phys. B }}
\def\epjc{{\em Eur. Phys. J. C }}
\def\plb{{\em Phys. Lett. B }}
\def\mpla{{\em Mod. Phys. Lett. A }}
\def\pr{{\em Phys. Rep. }}
\def\prv{{\em Phys. Rev. }}
\def\zpc{{\em Z. Phys. C }}
\def\zpa{{\em Z. Phys. A }}
\def\ppnp{{\em Prog. Part. Nucl. Phys. }}
\def\jpg{{\em J. Phys. G }}
\def\cpc{{\em Comput. Phys. Commun.}}
\def\app{{\em Acta Physica Pol. B }}
\def\aip{{\em AIP Conf. Proc. }}
\def\jhep{{\em J. High Energy Phy. }}
\def\ahep{{\em Adv. in High Eng. Phys. }}
\def\ijmpa{{\em Int. J. Mod. Phys. A }}
\def\psc{{\em Prog. Sci. Culture }}
\def\snc{{\em Suppl. Nuovo Cimento }}
\def\phy{{\em Physics }}
\def\arnps{{\em Annu. Rev. Nucl. Part. Sci. }}
\def\jpcs{{\em J. of Phys. Conf. Ser. }}
\def\ijmpe{{\em Int. J. Mod. Phys. E }}

\title{Energy and Centrality dependence of $dN_{\rm ch}/d\eta$ and $dE_{\rm T}/d\eta$ in Heavy-Ion Collisions from $\sqrt{s_{\rm NN}}$ =7.7 GeV to 5.02 TeV}
\author{Aditya Nath Mishra\inst{1}, Raghunath Sahoo\inst{1,}\thanks{\emph Corresponding Author: Raghunath.Sahoo@cern.ch}, Pragati Sahoo\inst{1}, Pooja Pareek\inst{1}, Nirbhay K. Behera\inst{1,}\thanks{\emph Current Affiliation: Department of Physics, College of Natural Science, Inha University, Incheon, South Korea- 402751}  \and 
Basanta K. Nandi\inst{2}
}                     
%
%
\institute{Discipline of Physics, School of Basic Sciences, Indian Institute of Technology Indore, Indore-453552, India \and 
Department of Physics, Indian Institute of Technology Bombay, Mumbai-400067, India}
\date{\today}
%
\abstract{
The centrality dependence of pseudorapidity density of charged
particles and transverse energy is studied for a wide range of
collision energies for heavy-ion collisions at midrapidity from 7.7
GeV to 5.02 TeV. A two-component model approach has been adopted to
quantify the soft and hard components of particle production, coming
from nucleon participants and binary nucleon-nucleon collisions,
respectively. Within experimental uncertainties, the hard component
contributing to the particle production has been found not to show any
clear collision energy dependence from RHIC to LHC.  The effect of centrality 
and collision energy in particle production seems to factor out with some degree of dependency
on the collision species. The collision of Uranium-like deformed nuclei opens up new challenges in understanding the energy-centrality factorization, which is evident from the centrality dependence of transverse energy density, when compared to collision of symmetric nuclei. 
} 
\authorrunning{Aditya Nath Mishra {\it et al.}}
\titlerunning{Energy and Centrality dependence...}
\maketitle
\section{INTRODUCTION}
\label{intro}
Under extreme conditions of high temperature and energy density, study of nuclear matter and especially the phase transition of hadronic matter to the Quark-Gluon Plasma (QGP) is one of the main goals of heavy-ion collision experiments, like RHIC at BNL, SPS and LHC at CERN. Envisaged by many theoretical works \cite{collins,bjorken,karsch}, coherent development of phenomenological
studies and experimental findings \cite{Arsene,Bback,AdamJ} support the fact of formation of a deconfined state of matter in the ultra-relativistic heavy-ion collisions. There are still both theoretical and experimental challenges to understand the properties of QGP. Among them, one of the most fundamental phenomena to be understood is the multiparticle production mechanism in heavy-ion collision experiments \cite{Mishra:2013dwa,Sarkisyan:2010kb,Sarkisyan:2016dzo,Sarkisyan:2015gca,Mishra:2014dta} and its relation with initial conditions. It has been proposed that two of the global observables such as pseudorapidity density of charged particles ($dN_{\rm ch}/d\eta$) and transverse energy density ($dE_{\rm T}/d\eta$) can be used as tools to understand the mechanism of particle production
\cite{bjorken,kataja}. The charged particle pseudorapidity density and
transverse energy are the first measurements as global observables of
matter created at extreme conditions of temperature and energy
density. Furthermore, $dN_{\rm ch}/dy$ and $dE_{\rm T}/dy$ are
related to the entropy and initial energy density of the system,
respectively \cite{bjorken,caruthers,wong}. 
The experimentally measured $dE_{\rm T}/d\eta$ is used for initial energy density calculation in Bjorken boost invariant hydrodynamics through a Jacobian transformation of rapidity to pseudorapidity.
%
Assuming complete chemical equilibrium of the medium, ideal hydrodynamics leads to total entropy conservation. In addition to that, in a simple case of boost invariant Bjorken expansion assuming ideal gas Equation of State (EoS), the entropy per unit rapidity and hence the rapidity density of particles are also conserved \cite{bjorken}.
So these two global observables can be used to provide constraints on the initial conditions \cite{wong,hijingPRL,ahepRev}. Several works  \cite{eskola,wangPRD,nayak1,nayak2,bhalerao} suggest that at the initial phase of collision, both perturbative and non-perturbative QCD processes take place. Non-perturbative processes involve soft gluon exchanges which come under soft processes and expected to scale with the number of participant nucleons. Hard scatterings occur (perturbative QCD) earlier to soft interactions and involve high-momentum transfer phenomena. Observables thus are expected to scale with number of binary nucleon-nucleon collisions \cite{hijingPRL}. To be specific, the processes involved for particle production at elementary level can be understood by disentangling the role of soft
process and hard scattering contributing to it
\cite{hijingPRD,kharjev}. It is reported in Ref,
\cite{hijingPRL,kharjev} that the charged particle pseudorapidity
distribution in nucleus-nucleus collisions (A+A) can be described by a
two-component model. In this work, using the same two-component model, we have tried to estimate the relative contribution of soft and hard processes in particle production by considering the centrality and collision energy dependence of $dN_{\rm ch}/d\eta$ and $dE_{\rm T}/d\eta$ for a wide range of experimental data starting from RHIC to LHC. 
\par
This paper is organized as follows. In section 2, we give a brief
introduction to the two-component model to estimate the hard and soft processes contribution to particle production. The details of the analysis methodology is presented in section 3. In section 4, we present the analysis results by fitting
two-component model to the experimental data of $dN_{\rm ch}/d\eta$
and $dE_{\rm T}/d\eta$ as complementary study by taking both the global observables. We discuss the energy-centrality factorization in section 5. The paper ends with summary and conclusion carrying some of the open issues in this direction in section 6.


\section{The Two-component model}
Nucleus-nucleus collision can be thought of superposition of many
nucleon-nucleon collisions. So to understand the mechanism of
multiparticle production, we need to investigate the processes at
fundamental level in nucleon-nucleon collisions. As we know, the major
contribution for particle production in nucleon-nucleon collisions is
from inelastic processes, which can again be divided into two parts: soft
non-perturbative processes and hard perturbative processes resulting into mini-jet and jet productions. Hence, the total inelastic cross section of the nucleon-nucleon collisions can be written as \cite{hijingPRD},

\begin{equation}
\sigma_{in}^{NN} = \int d^2b \left[ 1 - e ^{-\left( \sigma_{soft}(s) + \sigma_{jet}(s) \right) T_{NN}(b,s)} \right],
\label{inel}
\end{equation}
where $\sigma_{\rm soft}$ is the total cross section of the soft processes arising from the small momentum transfer to the constituent quarks as well as induced soft gluon radiation, $\sigma_{\rm jet}$ represents the integrated inclusive jet production cross section due to hard scattering of partons and can be calculated in the framework of perturbative QCD.  $T_{\rm NN}(b,s)$ is the partonic overlap function between the two nucleons at a given impact parameter $b$ and energy $s$. In the mean time, under the formalism of eikonal approach, the impact parameter representation of nucleon-nucleon collision can be written as \cite{wangPRD,WangpQCD},

\begin{equation}
\sigma_{in}^{NN} =  \pi \int_{0}^{\infty} d^2b \left(1 - e ^{-2\chi(b,s)}  \right) ,
\label{eikonal}
\end{equation}
where $\chi(b,s)$ is the eikonal function. Now comparing eq. (\ref{inel}) and  eq. (\ref{eikonal}), we have,

\begin{equation}
\chi(b,s) = \frac{1}{2} \sigma_{soft}(s)T_{NN}(b,s) + \frac{1}{2}\sigma_{jet}(s)T_{NN}(b,s).
\label{Ekrelation}
\end{equation}
Eq.(\ref{Ekrelation}) links the soft and hard processes with
geometrical quantities and the collision energy.  
Now using the eikonal approach, the number of jets produced in the nucleon-nucleon collisions can be calculated. To be noted here, in the string picture of Lund fragmentation model \cite{lund,hijingProg}, the wounded nucleons are treated as excited strings along the beam direction, suffer hard scattering and fragment to produce particles. So the hard scattering can be scaled with the number of binary nucleon-nucleon collisions ($N_{\rm coll}$) and the total numbers of jets in nuclear collisions can be calculated as,
\begin{equation}
N^{AA}_{jet} = T_{AA}(b)\sigma_{jet} ,
\label{njet}
\end{equation}
where $T_{\rm AA}(b)$ is the nuclear overlap function. In fact, the
right hand side of eq. (\ref{njet}) equals to $N_{\rm coll}$ according
to Glauber model \cite{Glauber}. After the hard scattering, the
remaining energy is used for soft interaction. It is observed from the
low energy $pp$ and heavy-ion collisions data that the $dN_{\rm
  ch}/d\eta$ normalised to number of participant nucleon ($N_{\rm
  part}$) pairs has a constant factor arising from soft process plus a
logarithmic energy dependent component. This soft process is
proportional to $N_{\rm part}$ \cite{hijingPRL}. Therefore, it is
assumed that all the participant nucleons contribute the same fraction
of energy to the soft process. Upto SPS energy, $dN_{\rm ch}/d\eta$
scales with $N_{\rm part}$. However, at RHIC and LHC energies, the
data show a monotonic rise with collision centrality
\cite{AliCent}. This is because, with increase of energy, the total
jet cross section increases much faster than the total inelastic cross
section resulting in the increase of number of minijets. Hence,
two-component model was introduced to describe the multiparticle
production at high energy nucleus-nucleus collisions, which includes
both the soft and hard processes contribution for particle production
\cite{hijingPRL}. In the framework of a two-component model, the $dN_{\rm ch}/d\eta$ for A+A collisions is given as,
\begin{equation}
\frac{dN_{ch}}{d\eta} = \langle N_{part} \rangle {\langle n \rangle}_{soft} + f \langle N_{coll} \rangle \frac{\sigma_{jet}^{AA}(s)}{\sigma_{in}^{NN}}.
\label{twocomp1}
\end{equation}

In eq. (\ref{twocomp1}), ${\langle n \rangle}_{soft}$ is the average multiplicity in the soft sector, the $\sigma_{\rm jet}^{\rm AA}(s)$ is the
average inclusive jet cross section per nucleon-nucleon collisions in
A+A collisions and $f$ is a constant factor. Now looking at
eq. (\ref{Ekrelation}) and the eikonal approach as given in
eq. (\ref{twocomp1}), the total multiplicity density in A+A collisions with
respect to multiplicity density of nucleon-nucleon collisions ($n_{\rm pp}$) can be rewritten in terms of contributions from soft and hard processes in a probabilistic way as \cite{kharjev},

\begin{equation}
\frac{dN_{ch}^{AA}}{d\eta} =  n_{pp} \left[ (1-x) \frac{ \langle N_{part} \rangle }{2} + x \langle N_{coll} \rangle \right] .
\label{twocompCh}
\end{equation}
In eq. (\ref{twocompCh}), $x$ represents the fraction of hard
processes and remaining (1-$x$) is the soft processes contributing to
the particle production. Transverse energy production in
nucleus-nucleus collision can also be treated on equal footing with
the charged particle production and hence, the two-component model for
transverse energy density can be written as follows \cite{ET-TwoComp},
\begin{equation}
\frac{dE_{T}^{AA}}{d\eta} =  \frac{dE_{T}^{pp}}{d\eta} \left[ (1-x) \frac{ \langle N_{part} \rangle }{2} + x \langle N_{coll} \rangle \right] .
\label{twocompEt}
\end{equation}
  
The geometrical quantities, like $N_{\rm part}$ and $N_{\rm coll}$,
can be estimated directly from the MC Glauber model for given
collision energy. The estimation of these quantities and then the relation between them are established in next section. Then the fraction of soft and hard processes are estimated by fitting two-component model to $dN_{\rm ch}/d\eta$ and $dE_{\rm T}/d\eta$ for a wide range of collision energies. 

\begin{figure}
\begin{center}
\resizebox{0.47\textwidth}{!}{%
  \includegraphics{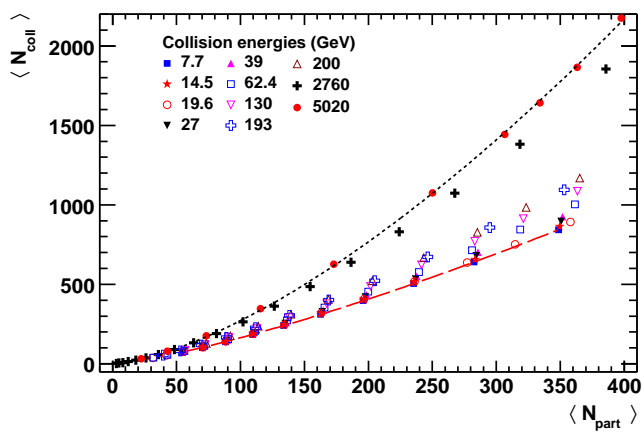}
}\caption{(Color online) The number of binary nucleon-nucleon collisions, $N_{\rm
    coll}$, as a function of number of participating nucleons,
  $N_{\rm part}$, which is a measure of collision centrality. The data
  are fitted with eq. (\ref{param}) at different
  collision energies from $\sqrt{s_{\rm NN}}$ = 7.7 GeV to 5.02 TeV.}
\label{fig:1}
\end{center}
\end{figure}


\begin{figure}
\begin{center}
\resizebox{0.47\textwidth}{!}{%
  \includegraphics{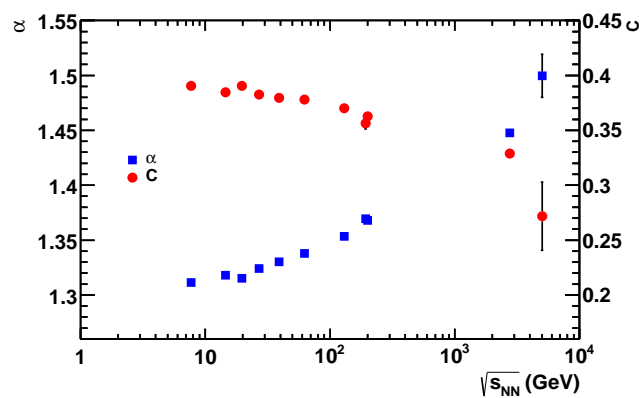}
}
\caption{(Color online) Parameter $\alpha$ and $C$, as a function of collision energy.} 
\label{fig:3}
\end{center}
\end{figure}


\section{Analysis Methodology}
According to Glauber model, for A+A collisions, $N_{\rm coll}$ scales as $A^{4/3}$ \cite{Glauber}, where $A$ is the mass number of the nucleus and is related to the nuclear radius as, $R=R_0~A^{\frac{1}{3}}$, with $R_0 = 1.2$ fm. Meanwhile, $N_{\rm part}$ directly depends on the interaction volume and again this volume is proportional to $A$ of a saturated nuclear density. Hence,  $N_{\rm coll} \propto N_{\rm part}^{4/3}$ and taking the geometry into consideration, $N_{\rm coll}$ can be parametrized in terms of $N_{\rm part}$ \cite{Glauber,phobosPRC70} as,

\begin{equation}
{ \langle N_{coll} \rangle =  C \times \langle N_{part}\rangle^\alpha },
\label{param}
\end{equation}
with $C$ and $\alpha$ constants. With the use of eq. (\ref{param}), eq. (\ref{twocompCh}) reduces to one variable function, i.e. $\langle N_{\rm part} \rangle$ and can be written as, 

\begin{equation}
\frac{dN_{ch}}{d\eta} = n_{pp}
\left[
(1-x)\frac{\langle N_{part} \rangle}{2} + x  C \langle N_{part} \rangle^\alpha %
\right] .
\label{oneComp}
\end{equation}
 
\begin{figure}
\begin{center}
\resizebox{0.47\textwidth}{!}{%
  \includegraphics{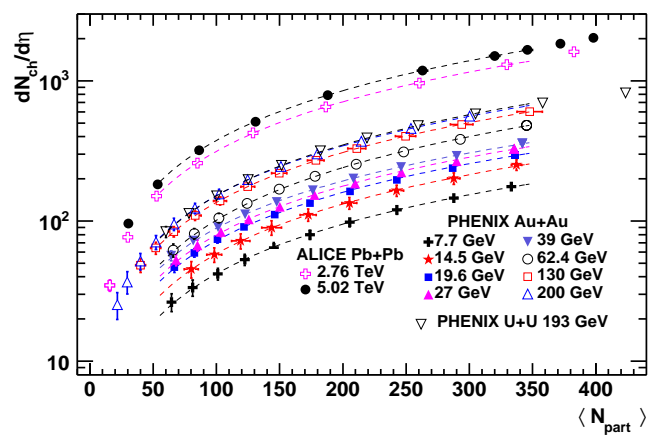}
}
\caption{ (Color online) Centrality dependence of $dN_{\rm ch}/d\eta$ at different
  RHIC energies (PHENIX experiment \cite{Adare:2015bua,Iordanova:2013jba})
  and the LHC energy (ALICE experiment
  \cite{Aamodt:2010cz,Adam:2015ptt}). Fitted lines are the corresponding two-component
  model fitting to extract the parameters $x$ and $n_{\rm pp}$. The fittings are performed in the same range of $N_{\rm part}$ for all energies.}
\label{fig:4}
\end{center}
\end{figure}

The values of $\langle N_{\rm coll}\rangle$ and $\langle N_{\rm
  part}\rangle$ are obtained by using nuclear overlap model
\cite{misko} which uses the Monte Carlo Glauber approach. Then the
obtained values of $\langle N_{\rm coll}\rangle$ are plotted against
$\langle N_{\rm part}\rangle$, which are shown in fig. \ref{fig:1}
and fitted by eq. (\ref{param}) to estimate the parameters $C$ and
$\alpha$. These parameters are estimated at different collision
energies spanning from 7.7 GeV to 5.02 TeV. The extracted values of
$C$ and $\alpha$ are shown as a function of $\sqrt{s_{\rm NN}}$ in fig.
\ref{fig:3}. While the parameter $C$ decreases with increase in
$\sqrt{s_{\rm NN}}$, the parameter, $\alpha$, in eq. (\ref{param})
increases monotonically with $\sqrt{s_{\rm NN}}$. These parameters are further used as inputs for eq. (\ref{oneComp}). Then using eq. (\ref{oneComp}), the values of $x$ and $n_{\rm pp}$ are obtained for various centrality data at different collision energies. The experimental data used in this paper for $dN_{\rm ch}/d\eta$ as a function of $N_{\rm part}$ for Au+Au collisions at $\sqrt{s_{\rm NN}}$ = 7.7, 14.5, 19.6, 27, 39, 62.4, 130 and 200 GeV, and for U+U collisions at $\sqrt{s_{\rm NN}}$ = 193 GeV, are taken from PHENIX experiment \cite{Adare:2015bua,Iordanova:2013jba,Adler:2004zn}. Data for Pb+Pb collisions at $\sqrt{s_{\rm NN}}$ = 2.76 \cite{Aamodt:2010cz} and 5.02 TeV \cite{Adam:2015ptt} are taken from ALICE experiment at the LHC. The two-component model fittings are shown in fig. \ref{fig:4}. All experimental data given here are already corrected for $p_{\rm T} =0$ and for consistency, we fit all the data sets in the same range of $N_{\rm part}$.

\begin{figure}
\begin{center} 
\resizebox{0.47\textwidth}{!}{%
  \includegraphics{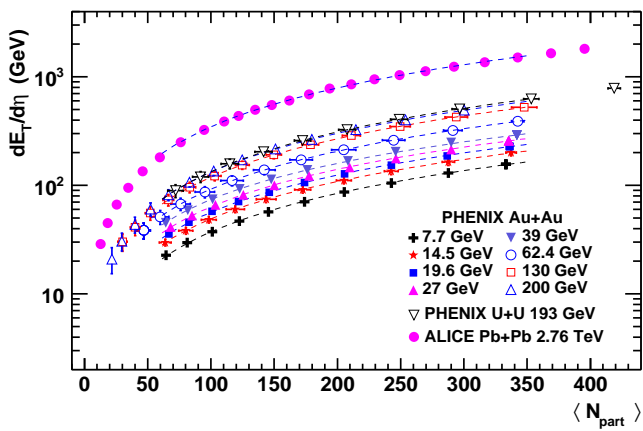}
}
\caption{(Color online) Centrality dependence of $dE_{\rm T}/d\eta$ at different RHIC
  energies (taken by PHENIX experiment \cite{Adare:2015bua,Iordanova:2013jba}) and for LHC (by ALICE experiment \cite{Adam:2016thv}). Fitted lines are the corresponding two-component model fittings, to extract the parameters $x$ and $dE_{\rm T}^{pp}/d\eta$.
  The fittings are performed in the same range of $N_{\rm part}$ for all energies.} 
\label{fig:6}
\end{center}
\end{figure}

We repeat the similar exercise taking the centrality data for transverse energy densities of produced particles, $dE_{\rm T}/d\eta$ at different $\sqrt{s_{\rm NN}}$. To obtain the values of $x$ and $dE_{\rm T}^{pp}/d\eta$, we use the same form of eq. (\ref{oneComp}) only by replacing $dN_{\rm ch}/d\eta$ with $dE_{\rm T}/d\eta$ as follows.
 \begin{equation}
\frac{dE_{T}}{d\eta} = \frac{dE_{T}^{pp}}{d\eta}  \left[
(1-x)\frac{\langle N_{part} \rangle}{2} + x  C \langle N_{part} \rangle ^\alpha %
\right] .
\label{oneCompEt}
\end{equation}
 For this analysis, we use the data taken by the PHENIX
 experiment for Au+Au collisions at $\sqrt{s_{\rm NN}}$ = 7.7, 14.5, 19.6, 27, 39, 62.4, 130 and 200
 GeV \cite{Adare:2015bua,Iordanova:2013jba}. The corresponding LHC data for Pb+Pb collisions at $\sqrt{s_{\rm NN}}$ =
 2.76 TeV are taken from the ALICE experiment \cite{Adam:2016thv}. The data are fitted by eq. (\ref{oneCompEt}) and are shown in fig. \ref{fig:6}. The detailed results are discussed in the following section. The errors shown in the figures are the quadratic sum of systematic and statistical errors on experimental data. In case of the extracted parameters, these are from the phenomenological fittings. Where not visible, these are within the marker size.


\section{Results}
 
\begin{figure}
\begin{center}
\resizebox{0.47\textwidth}{!}{%
  \includegraphics{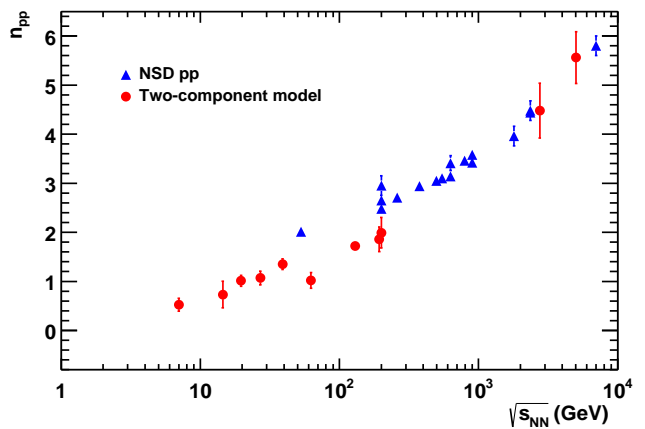}
}
\caption{ (Color online) Charged particle multiplicity density in $pp$ collision, $n_{\rm
    pp}$, as a function of center of mass energy. The solid circles
  show the values of $n_{\rm pp}$, estimated from heavy-ion data using
  eq.(\ref{oneComp}) (fittings shown in fig. \ref{fig:4}), the solid
 triangles show the values of $n_{\rm pp}$ measured in non-single
  diffractive (NSD) $pp$ collisions by ALICE \cite{kAamodt} and CMS \cite{Khachatryan} experiments at LHC, and from $\bar{p}p$ collisions by UA5 at CERN SPS \cite{Alner,gAlner} and ISR ($\sqrt{s_{\rm pp}}$ = 53 GeV), by CDF Collab. at Fermilab \cite{Abe}.}  
\label{fig:5}
\end{center}
\end{figure}

\begin{figure}
\begin{center}
\resizebox{0.47\textwidth}{!}{%
  \includegraphics{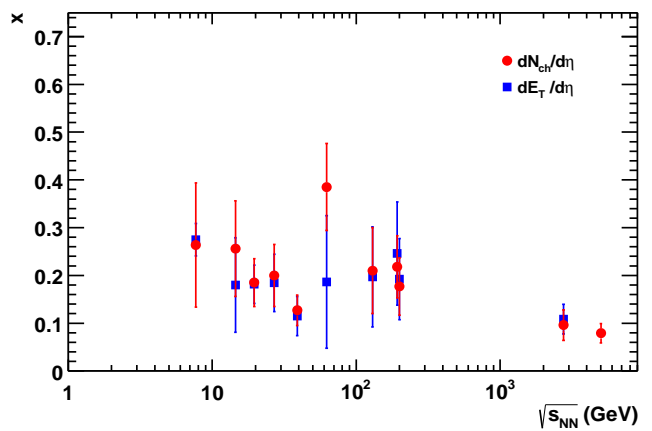}
}
\caption{(Color online) The hard scattering component, $x$, as a function of collision energy, which seems not to show any clear collision energy dependent behavior in heavy-ion collision environment.}
\label{fig:7}
\end{center}
\end{figure}

\begin{figure}
\begin{center}
\resizebox{0.47\textwidth}{!}{%
  \includegraphics{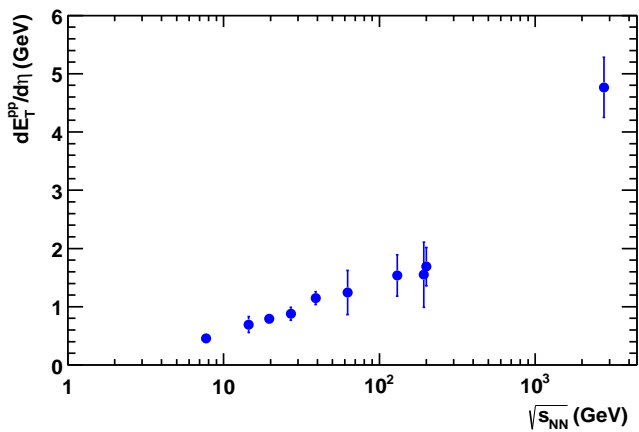}
}
\caption{(Color online) $dE_{\rm T}^{pp}/d\eta$ as a function of center of mass
  energy, extracted from two-component model fitting in centrality
  data of $dE_{\rm T}/d\eta$. Fittings are shown in fig. \ref{fig:6}
  for different energies.} 
\label{fig:8}
\end{center}
\end{figure}

The above discussed two-component approach is used to analyze the centrality data for $dN_{\rm ch}/d\eta$ and $dE_{\rm T}/d\eta$ for various energies. Figure \ref{fig:4} shows the centrality dependence of $dN_{\rm
  ch}/d\eta$ at midrapidity for different collision energies ranging
from 7.7 GeV to 5.02 TeV. The extracted values of $n_{\rm pp}$ are
shown as a function of $\sqrt{s_{\rm NN}}$ in fig. \ref{fig:5}. In
the same figure, we have also shown the experimentally measured values
of $n_{\rm pp}$, where the values of $n_{\rm pp}$ measured in
non-single diffractive (NSD) $pp$ collisions are from ALICE
\cite{kAamodt} and CMS experiments \cite{Khachatryan} at LHC, and in
NSD $\bar{p}p$ collisions from UA5 at CERN SPS \cite{Alner,gAlner}
and ISR ($\sqrt{s_{\rm pp}}$ = 53 GeV), by CDF collaboration at
Fermilab \cite{Abe}. Our estimated values are in good agreement with
the experimental measurements. The extracted hard scattering parameter
$x$ is shown in fig. \ref{fig:7} for $dN_{\rm ch}/d\eta$. As a
complementary measurement, adopting exactly the same method as above,
the values of $x$ obtained from two-component fit to the $dE_{\rm
  T}/d\eta$ centrality data at various collision energies are also shown in
fig. \ref{fig:7}. The values of $x$ both for $dN_{\rm ch}/d\eta$ and
$dE_{\rm T}/d\eta$ are consistent for different collision energies
within the experimental uncertainties. The PHOBOS collaboration has
also done similar analysis for $dN_{\rm ch}/d\eta$ at $\sqrt{s_{\rm
    NN}}$ = 19.6, 130 and 200 GeV. The observed value of the fraction of
hard interactions, $x$ = 0.13 $\pm$ 0.01 (stat) $\pm$ 0.05
(syst) at the above energies  \cite{Alver,BACKB}. We extend this analysis to include the RHIC Beam Energy Scan (BES) and the LHC measurements. In Table \ref{tab:table2}, the values of average multiplicity density
in $pp$ collision, $n_{\rm pp}$, and hard scattering component $x$ are
given, which are extracted from the two-component model fit to charged
particle multiplicity density, $dN_{\rm ch}/d\eta$ data (shown in fig.
\ref{fig:4}). The charged particle multiplicity density in $pp$ collisions, $n_{\rm pp}$, extracted by fitting the two-component model to $dN_{\rm ch}/d\eta$ centrality data, shows a monotonic increase with increasing
$\sqrt{s_{\rm NN}}$. This is in agreement with the experimental
observations, as could be seen from fig. \ref{fig:5}. From fig.
\ref{fig:8}, it can be observed that $dE_{\rm T}^{pp}/d\eta$ also
shows a similar behavior like $n_{\rm pp}$, {\it i.e.} shows a monotonic rise with increasing $\sqrt{s_{\rm NN}}$.
The hard scattering component, $x$, seems to be independent of collision energy, within systematic uncertainties. This is observed for both the global observables. Although similar behaviour is observed in PHOBOS analysis, our numbers are different because of different data sets, where different low-$p_{\rm T}$ cut-off are used.

\begin{table}[h]
\caption{\label{tab:table2} Hard scattering component, $x$, and the charged particle multiplicity in pp collisions, $n_{\rm pp}$ , at different centre-of-mass energies, as obtained from the fittings in fig. 3}
\begin{tabular}{|c|c|c|c|}
\hline
& $\sqrt{s_{\rm NN}}$ (GeV) & $x$& $n_{\rm pp}$\\ \hline
Au $+$ Au & 7.7 & 0.264 $\pm$ 0.130 & 0.524 $\pm$ 0.132\\
Au $+$ Au & 14.5 & 0.256 $\pm$ 0.010 & 0.731 $\pm$ 0.269\\
Au $+$ Au & 19.6 & 0.185 $\pm$ 0.050 & 1.012 $\pm$ 0.111\\
Au $+$ Au & 27 & 0.200 $\pm$ 0.065 & 1.069 $\pm$ 0.140\\
Au $+$ Au & 39 & 0.127 $\pm$ 0.032 & 1.353 $\pm$ 0.107\\   
Au $+$ Au & 62.4 & 0.385 $\pm$ 0.091 & 1.022 $\pm$ 0.160 \\
Au $+$ Au & 130  & 0.210 $\pm$ 0.090 & 1.720 $\pm$ 0.089 \\
U $+$ U & 193  & 0.218 $\pm$ 0.065 & 1.864 $\pm$ 0.251 \\
Au $+$ Au & 200  & 0.177 $\pm$ 0.060 & 1.991 $\pm$ 0.314 \\
Pb $+$ Pb & 2760 & 0.096 $\pm$ 0.032 & 4.480 $\pm$ 0.560 \\
Pb $+$ Pb & 5020 & 0.079 $\pm$ 0.020 & 5.562 $\pm$ 0.533 \\   
\hline
\end{tabular}
\end{table}

\begin{figure}
\begin{center}
\resizebox{0.47\textwidth}{!}{%
  \includegraphics{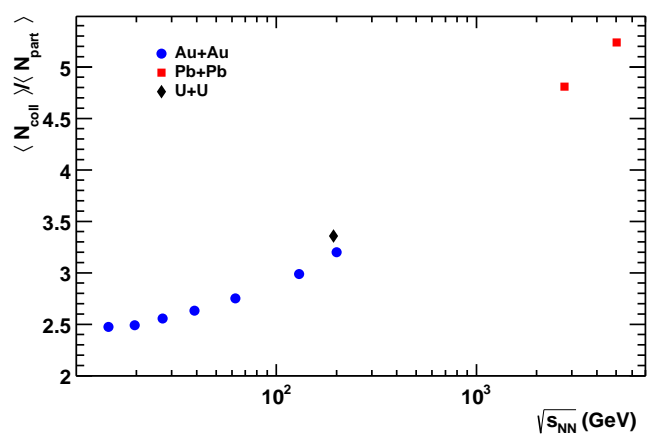}
}
\caption{(Color online) $N_{\rm coll}$ scaled with $N_{\rm part}$ as a function of
  collision energy for most central collisions. Both $\left<N_{\rm coll}\right>$ and $\left<N_{\rm part}\right>$ are obtained using nuclear overlap model, which uses a Glauber Monte Carlo approach.} 
\label{fig:9}
\end{center}
\end{figure}

In fig. \ref{fig:9}, $\left<N_{\rm coll}\right>$ scaled with $\left<N_{\rm part}\right>$ as a
function of $\sqrt{s_{\rm NN}}$ at different center-of-mass energies
shows a monotonic increase. This shows an expected increase of hard scattering, which scales with $\left<N_{\rm coll}\right>$, as a function of collision energy. U+U collisions at $\sqrt{s_{\rm NN}}$ =193 GeV
shows a higher value of $\left<N_{\rm coll}\right>$/ $\left<N_{\rm part}\right>$ compared to Au+Au
collisions at $\sqrt{s_{\rm NN}}$ =200 GeV. This is because of higher mass
number for Uranium. The observation of increase in the above ratio is expected because when the center-of-mass energy increases the number of binary nucleon-nucleon
collisions increases faster than the number of participants. The hard
scattering contribution in particle production is expected to increase
with increasing $\sqrt{s_{\rm NN}}$. But we don't observe that in
heavy-ion collisions. A similar analysis for $dN_{\rm ch}/d\eta$ has
been done in Ref.\cite{stephans} for few energies. Our results for
both $dN_{\rm ch}/d\eta$ and $dE_{\rm T}/d\eta$ are consistent with
the above findings. However, this observation doesn't go with 
the theoretical expectations \cite{hijingPRL,kharjev}. The following factors may be responsible for the above observations: (i) medium effect- the loss of information on contribution of hard scattering to particle production because of the suppression of high-$p_{\rm T}$ hadrons and jets  \cite{Adams,Jadams} in the dense medium created in heavy-ion collision, (ii) interference of minijets- with increase of collision energy, the number of minijets increases. So particle production from minijets dilutes the significance of contribution of hard process with increasing energy.


\section{Energy-Centrality Factorization}
In Ref. \cite{Alver}, first time it was demonstrated that the normalized charged particle production per participant pair at midrapidity can be factorized in terms of collision energy and collision centrality as follows

\begin{equation}
\frac{1}{0.5N_{part}} \frac{dN_{ch}}{d\eta} = f\left(s\right)g\left(N_{part}\right),
\label{eqn:5}
\end{equation}
where $f(s)$ and $g(N_{\rm part})$ factor out the collision energy ($s$) and centrality dependence, respectively. The parametric form of $f\left(s\right)$ and $g\left(N_{\rm part}\right)$  for Au+Au collisions are found to be as follows \cite{Alver}.

\begin{equation}
f\left(s\right) = 0.0147 \left[ \ln (s) \right]^2 + 0.6 ,
\label{eqn:6}
\end{equation}

\begin{equation}
g\left(N_{part}\right) = 1 +  0.095N_{part}^{1/3}.
\label{eqn:7}
\end{equation}

It was also observed that for Cu+Cu collisions both the coefficients of $f\left(s\right)$ remain the same, however, one of the coefficients of $g\left(N_{\rm part}\right)$ changes as follows

\begin{equation}
g\left(N_{part}\right) = 1 +  0.129N_{part}^{1/3}.
\label{eqn:8}
\end{equation}

Change in coefficient of $g\left(N_{\rm part}\right)$ while going from Au+Au collision to Cu+Cu collision data represents the collision species dependence of $g\left(N_{\rm part}\right)$. We have extended the above discussed idea of energy-centrality factorization in the multiparticle production processes spanning an energy domain of few GeV to TeV energies taking $dN_{\rm ch}/{d\eta}$ and $dE_{\rm T}/{d\eta}$, as complementary global observables. This factorization essentially deals with collision energy and collision geometry, or directly the number of participant contribution to particle production. The question here is- do both of them contribute independently to multiparticle production processes? The answer to this question will be evident, if one studies the factorization concept for a large energy range taking different collision species. Here, we have taken the centrality data for both the observables from 7.7 GeV beam energy scan (BES) at RHIC to the available top energy at LHC, 5.02 TeV and different collision species like- Au+Au, U+U and Pb+Pb. 

\begin{figure}
\begin{center}
\resizebox{0.47\textwidth}{!}{%
  \includegraphics{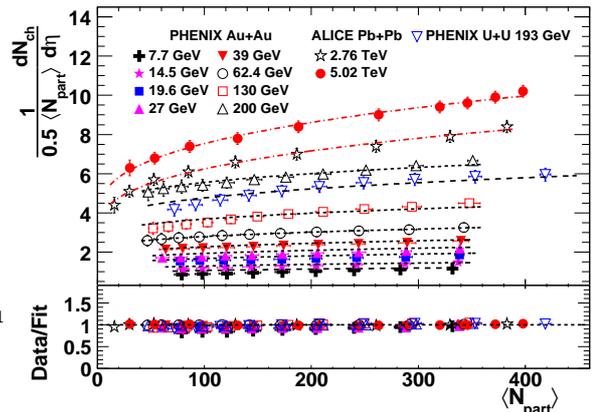}
}
\caption{(Color online) Upper panel: Centrality dependance of pseudorapidity density of charge particle multiplicity at midrapidity per participant pair, $dN_{\rm ch}/d\eta/\langle N_{\rm part}/2 \rangle$. Eq. (\ref{eqn:5}) is fitted to both RHIC (dashed line) and LHC data (solid and dashed dotted line). Lower panel: The ratio of data and fitting indicating goodness of the fit.}
\label{fig:Fac1}
\end{center}
\end{figure}

For Au+Au collisions at RHIC BES from 7.7 to 200 GeV, we observe similar results within fitting errors as is observed by the PHOBOS experiment \cite{Alver}. The fitted functions which are depicted in fig.~\ref{fig:Fac1} are given by
\begin{equation}
f\left(s\right) = (0.0147 \pm 0.0006) \left[ \ln (s) \right]^2 + (0.601\pm 0.030) ,
\label{eqn:9}
\end{equation}

\begin{equation}
g\left(N_{part}\right) = (1.001\pm 0.042) +  (0.0955 \pm 0.035) N_{part}^{1/3}.
\label{eqn:10}
\end{equation}
For U+U collisions at 193 GeV, we observe that only the second coefficient of $g\left(N_{part}\right)$ changes to $0.113 \pm 0.005$, which reflects the effect of collision geometry. To study the factorization at LHC energies we fit the same function 
to Pb+Pb data at $\sqrt{s_{\rm NN}}$ = 2.76 and 5.02 TeV, taken by ALICE detector at LHC \cite{Aamodt:2010cz,Adam:2015ptt} and found that the coefficients of $f\left(s\right)$ do not change, however, both the coefficients of $g\left(N_{\rm part}\right)$ change and are given by

\begin{equation}
g\left(N_{part}\right) = (0.696 \pm 0.09) +  (0.184 \pm 0.015) N_{part}^{1/3}.
\label{eqn:11}
\end{equation}
These results are shown in the upper panel of fig.~\ref{fig:Fac1}. In the lower panel, the ratio of data and fitted factorization function are shown, which show a very good agreement towards energy-centrality factorization in a broad range of collision energy.

\begin{figure}
\begin{center}
\resizebox{0.47\textwidth}{!}{%
  \includegraphics{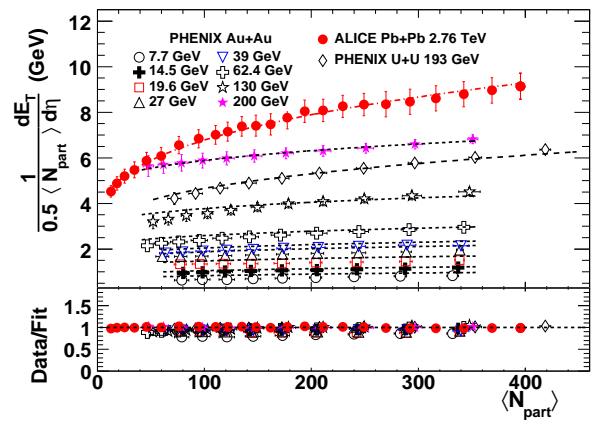}
}
\caption{(Color online) Upper panel: Centrality dependence of pseudorapidity density of transverse energy at midrapidity normalized to $\langle N_{\rm part} \rangle$ pair, $dE_{\rm T}/d\eta/\langle N_{\rm part}/2 \rangle$ at different collision energies. Eq. (\ref{eqn:5}) is fitted to RHIC and LHC data, by replacing $N_{\rm ch}$ by $E_{\rm T}$ in the equation.
Lower panel: The ratio of data and fitting indicating goodness of the fit.}
\label{FactorEt}
\end{center}
\end{figure}

For a complementarity study, we do a similar analysis for the other global observable, $dE_{\rm T}/d\eta$ from 7.7 GeV to 2.76 TeV for Au+Au, U+U and Pb+Pb collisions. This is shown in the upper panel of fig.~\ref{FactorEt}. The lower panel shows 
the ratio of data and fitting of the factorization function representing the goodness of the fits. For Au+Au collisions at the discussed centre of mass energies, we get the following factorization functions.

\begin{equation}
f\left(s\right) = (0.0147 \pm 0.0006) \left[ \ln (s) \right]^2 + (0.4709\pm 0.030) ,
\label{eqn:12}
\end{equation}

\begin{equation}
g\left(N_{part}\right) = (1.011\pm 0.05) +  (0.083 \pm 0.004) N_{part}^{1/3}.
\label{eqn:13}
\end{equation}

However, as expected, for Pb+Pb collisions, eq.~(\ref{eqn:12}) remains the same and the effect of change of geometry is reflected in $g\left(N_{part}\right)$ as:

\begin{equation}
g\left(N_{part}\right) = (0.592\pm 0.122) +  (0.2199 \pm 0.04) N_{part}^{1/3}.
\label{eqn:14}
\end{equation}

Unlike Au+Au and Pb+Pb collisions, U+U collisions are different because of its deformed nature of the nuclei, where the 
effect of flow is correlated with the initial collision geometry. Transverse energy is affected by radial flow and hence the initial collision geometry affects the final state $dE_{\rm T}/d\eta$. One expects stronger radial flow due to higher medium density, as is the case of U+U collisions \cite{Heinz-PRL94}. We would therefore expect different energy-centrality factorization behaviour for U+U collisions in case of  $dE_{\rm T}/d\eta$. In this case, when the factorization function is fitted to the centrality-dependent number-of-participants-normalized $dE_{\rm T}/d\eta$, eq.~(\ref{eqn:12}) remains the same, whereas we get the following function for $g\left(N_{part}\right)$.

\begin{equation}
g\left(N_{part}\right) = (0.6438\pm 0.04) +  (0.1523 \pm 0.003) N_{part}^{1/3}.
\label{eqn:15}
\end{equation}
Unlike $N_{\rm coll}$, which increases almost $\sim 50\%$ from RHIC $\sqrt{s_{\rm NN}}$ = 200 GeV to LHC $\sqrt{s_{\rm NN}}$ = 2.76 TeV for a similar collision species, ({\it e.g.} Au+Au), $N_{\rm part}$ increase is around $2.5\%$. This weak dependency of $N_{\rm part}$ on collision energy leads to a possible factorization behaviour in heavy-ion collisions.
The collision energy-centrality factorization is an interesting observation in heavy-ion collisions, which is like separating the variable-dependent wave functions in a quantum mechanical many-body system with symmetry.

 

\section{SUMMARY AND CONCLUSIONS}
The centrality dependence of the global observables
like pseudorapidity distribution of charged particles and transverse
energy density can be explained by a two-component model, which
accounts for the contribution of soft processes and hard scattering at
the initial stage of the collisions. Theoretical model studies suggest that with increase of collision
energy there is an increase of hard scattering. We study the two-component model
to investigate the probability of hard scattering contribution for
particle production as a function of collision energy. Through a
complementary measurement of the hard scattering component, $x$, in
$dN_{\rm ch}/d\eta$ and $dE_{\rm T}/d\eta$, we observe $x$ not to 
show any clear collision energy dependence from few GeV to TeV energies. It is observed that the coefficient of the hard-scattering,
{\it i.e.} $x$, does not increase with collision energy, whereas the
fraction of $N_{\rm{coll}}/N_{\rm{part}}$ shows a monotonic increase. In
addition, the QCD cross-sections also increase with energy. This gives
the evidence that the used two-component model fails to bring out the
information about the hard-scattering contribution to particle
production in a heavy-ion collision environment, where the medium effects play a vital role. This goes in line
with the observation of suppression of high-$p_{\rm T}$ hadrons and
jets in the dense medium created in heavy-ion collisions
\cite{Arsene,Bback,AdamJ}.  This leaves out an open question to devise a method to study the relative contribution of hard and soft processes towards the particle production in heavy-ion collisions. There are some theoretical works which suggest that the two-component model can be written as the sum of soft processes and the minijet cross sections, which need further investigations \cite{minijetTwocomp}. The discussed two-component model appears to be very crude and empirical in nature while bringing out the desired information on soft and hard scattering contribution to multiparticle production in heavy-ion collisions. The centrality and energy dependence of charged particle and transverse energy production seems to factor out with some degree of dependency on the collision species, which goes inline with the earlier observations \cite{Alver}. However, the collisions of U+U would be interesting to study in details because of the effect of the initial geometry.
\\
\\

\end{document}